\documentclass[sigconf, anonymous=False]{acmart}
\usepackage{bm}
\usepackage{amsmath}
\usepackage{multirow}
\usepackage{natbib}
\usepackage{makecell}
\usepackage{graphicx}
\usepackage{etoolbox}

\makeatletter
\g@addto@macro\bfseries{\boldmath}
\makeatother

\AtBeginDocument{%
  \providecommand\BibTeX{{%
    \normalfont B\kern-0.5em{\scshape i\kern-0.25em b}\kern-0.8em\TeX}}}

\setcopyright{acmcopyright}
\copyrightyear{2021} 
\acmYear{2021} 
\setcopyright{rightsretained} 
\acmConference[SIGIR '21]{Proceedings of the 44th International ACM SIGIR Conference on Research and Development in Information Retrieval}{July 11--15, 2021}{Virtual Event, Canada}
\acmBooktitle{Proceedings of the 44th International ACM SIGIR Conference on Research and Development in Information Retrieval (SIGIR '21), July 11--15, 2021, Virtual Event, Canada}
\acmDOI{10.1145/3404835.3462973}
\acmISBN{978-1-4503-8037-9/21/07}

\begin{document}
\fancyhead{}
\sloppy
\title{New Insights into Metric Optimization\\for Ranking-based Recommendation}

\author{Roger Zhe Li}
\affiliation{%
  \institution{Delft University of Technology}
  \city{Delft}
  \country{The Netherlands}
}
\email{Z.Li-9@tudelft.nl}

\author{Juli\'an Urbano}
\affiliation{%
  \institution{Delft University of Technology}
  \city{Delft}
  \country{The Netherlands}
}
\email{J.Urbano@tudelft.nl}

\author{Alan Hanjalic}
\affiliation{%
  \institution{Delft University of Technology}
  \city{Delft}
  \country{The Netherlands}
}
\email{A.Hanjalic@tudelft.nl}

\renewcommand{\shorttitle}{New Insights into Metric Optimization for Ranking-based Recommendation}

\begin{abstract}
Direct optimization of IR metrics has often been adopted as an approach to devise and develop ranking-based recommender systems. Most methods following this approach (e.g. TFMAP, CLiMF, Top-N-Rank) aim at optimizing the same metric being used for evaluation, under the assumption that this will lead to the best performance. A number of studies of this practice bring this assumption, however, into question.
In this paper, we dig deeper into this issue in order to learn more about the effects of the choice of the metric to optimize on the performance of a ranking-based recommender system. We present an extensive experimental study conducted on different datasets in both pairwise and listwise learning-to-rank (LTR) scenarios, to compare the relative merit of four popular IR metrics, namely $RR$, $AP$, $nDCG$ and $RBP$, when used for optimization and assessment of recommender systems in various combinations.
For the first three, we follow the practice of loss function formulation available in literature. For the fourth one, we propose novel loss functions inspired by $RBP$ for both the pairwise and listwise scenario.
Our results confirm that the best performance is indeed not necessarily achieved when optimizing the same metric being used for evaluation. In fact, we find that $RBP$-inspired losses perform at least as well as other metrics in a consistent way, and offer clear benefits in several cases. Interesting to see is that $RBP$-inspired losses, while improving the recommendation performance for all uses, may lead to an individual performance gain that is correlated with the activity level of a user in interacting with items. The more active the users, the more they benefit. Overall, our results challenge the assumption behind the current research practice of optimizing and evaluating the same metric, and point to $RBP$-based optimization instead as a promising alternative when learning to rank in the recommendation context.
\end{abstract}



\maketitle

\section{Introduction}\label{sec:intro}


Offline evaluation of ranking-based recommender systems generally relies on effectiveness metrics from Information Retrieval (IR). These metrics quantify the quality of a ranked list of items in terms of their relevance \citep{DBLP:journals/ir/ValcarceBPC20} and according to the particular evaluation criteria they capture. Therefore, it has been seen as an intuitive and logical choice to learn a ranking model for a recommender system by directly optimizing the metric used for evaluation \citep{DBLP:conf/sigir/Liu10}. Different ranking approaches have been designed along this line, aiming at achieving a better recommendation performance~\citep{DBLP:conf/recsys/ShiLH10,DBLP:conf/sigir/ShiKBLHO12,DBLP:conf/recsys/ShiKBLOH12,DBLP:conf/bigdataconf/LiangHDH18}. 

Previous research shows, however, that optimizing the metric used for evaluation is not necessarily the best approach. Different IR metrics reflect different aspects of retrieval performance \citep{DBLP:conf/sigir/AmigoGV13, DBLP:conf/sigir/JonesTSS15, DBLP:journals/ir/SakaiK08,DBLP:conf/airs/Moffat13}, and do so to different extents. Although no metric covers all evaluation criteria, there is evidence that some metrics are more informative than others~\citep{DBLP:conf/sigir/AslamYP05, DBLP:journals/ir/YilmazR10, DBLP:conf/www/AshkanC11}.
This may enable them to, when used for optimization, achieve the best performance in view of a given evaluation metric other than itself, or even in view of multiple target evaluation criteria. Empirical results show that the advantage in informativeness can indeed be transferred into higher effectiveness when these metrics are used for optimization. Results on web search \citep{DBLP:conf/sigir/DonmezSB09} and text retrieval \citep{DBLP:journals/ir/YilmazR10} show that, when targeting at less informative metrics such as $P@10$ and $RR$, optimizing more informative metrics, like $AP$ or $nDCG$, can perform even better than optimizing the less informative metrics themselves.

The discussion above points to the possibility to achieve better ranking performance according to the evaluation criterion captured by metric $X$ if we do not optimize for $X$, but for a more informative metric $Y$ instead. 
This, however, is more likely to be successful when the metrics $X$ an $Y$ are highly correlated. From the perspective of ranking effectiveness, correlation would typically arise between metrics that share characteristics, such as top-weightedness and localization. According to~\citet{DBLP:conf/airs/Moffat13}, correlation in view of such properties could tell something about the ability of the metrics to reveal the same aspects of the system behavior, as opposed to non-correlated metrics that reveal different aspects of the system behavior. 
Previous research on text retrieval and web search has shown high level of correlation among many metrics, such as between $nDCG$ and $AP$, $RR$ and $AP$, or $P@k$ and $AP$~\cite{DBLP:conf/sigir/WebberMZS08,  DBLP:conf/ecir/GuptaKKL19,DBLP:journals/kais/BacciniDLM12}.
When targeting at maximizing one specific evaluation metric, one should therefore search among informative metrics correlated to the evaluation target to find the most effective optimization objective.
Motivated by the above, in this paper we revisit the choice of the metric to optimize in a ranking-based recommendation scenario. Our goal is to provide more insights into this optimization scenario, broadening the possibilities to benefit from such a metric choice, compared to what is commonly reported in the literature. 

Although several popular metrics like $RR$, $nDCG$ and $AP$ have been applied to the task, $RBP$ (Rank-Biased Precision)~\citep{DBLP:journals/tois/MoffatZ08}, another important effectiveness metric widely used in traditional IR tasks, has hardly been applied for training and/or testing recommendation models. Nonetheless, $RBP$ is an informative \citep{DBLP:conf/airs/Moffat13, DBLP:journals/tois/MoffatBST17, DBLP:conf/sigir/AmigoSA18} and flexible metric that incorporates a simple user model through a \emph{persistence} parameter. Since varying the value of the persistence parameter makes $RBP$ correlated with different groups of metrics~\citep{DBLP:journals/ir/SakaiK08, DBLP:conf/adcs/MoffatST12}, 
$RBP$ has the potential to optimize for different metrics that reflect different evaluation criteria~\cite{DBLP:journals/tois/MoffatZ08, DBLP:conf/ecir/GuptaKKL19}. 

Following the spirit of existing metric optimization processes, we propose novel objective functions inspired by $RBP$ for both the pairwise and listwise learning-to-rank (LTR) paradigms. In this way, we enable $RBP$ to join the other metrics and serve as both the learning objective and evaluation target for our investigation. Specifically for the listwise case, we will show that minimizing the proposed $RBP$-based objective function provides an elegant instrument to directly optimize for the ranking positions of the relevant items.
Furthermore, the proposed $RBP$-based listwise loss function is independent of the persistence parameter $p$, which makes it possible to conduct $RBP$-based optimization using a single unified framework, regardless of the target persistence to evaluate.

Empirical results obtained on four real-world datasets point to the following main insights:
\begin{enumerate}
    \item The assumption behind the practice to optimize and evaluate ranking-based recommender systems using the same metric does not necessarily lead to the best performance.
    \item $RBP$-inspired losses perform at least as well as other metrics in a consistent way, and offer clear benefits in several cases. This makes $RBP$ a promising alternative when learning to rank in the recommendation context. 
    \item $RBP$-based listwise optimization leads ---relative to other metrics--- to a significantly better ranking performance for active users with more of the relevant interactions, compared to less active users. However, this performance bias does \emph{not} come at the cost of reducing utility for inactive users. On the contrary, the ranking performance improves for \emph{all} users, only to different degrees.
\end{enumerate}

The remainder of this paper is organized as follows. In Section~\ref{sec:related}, we position our contribution in the context of the related previous work. Section~\ref{sec:method} describes the technical details of the models we use for a direct optimization of IR metrics. Section~\ref{sec:exp} introduces the design and protocols of our experiments, the results of which we present and discuss in Section~\ref{sec:res}. Section~\ref{sec:conclusion} concludes the paper with pointers to future work.

\section{Related Work}\label{sec:related}
Direct optimization of IR metrics is a logical way of building ranking-based recommenders. 
Despite the fact that almost any IR metric can be transformed into an objective function for optimization, the choice of metric to optimize for maximizing the ranking effectiveness remains non-trivial. Intuitively, choosing to optimize more informative metrics helps with achieving higher ranking effectiveness. Several studies \citep{DBLP:conf/sigir/AslamYP05, DBLP:journals/ir/YilmazR10, DBLP:conf/www/AshkanC11}, inspired by \citet{shannon2001mathematical} and \citet{jaynes1982rationale}, have assessed the informativeness of IR metrics by how well they constrain a maximum entropy distribution over the relevance of ranked items, in the sense that such distribution accurately estimates the precision-recall curve or other metric scores for the same items. The results indicate that more complex metrics, such as $nDCG$ and $AP$, are more informative than simpler metrics~\cite{DBLP:conf/sigir/WebberMZS08}. 

One way of achieving optimal ranking with respect to a given metric is to deploy a \emph{pairwise} LTR approach. The pairwise LTR paradigm considers relevant-irrelevant (positive-negative) item pairs, and aims at maximizing the change in the considered IR metric caused by a ranking position swap. This idea lays the foundation for a batch of models, including LambdaRank~\citep{DBLP:conf/nips/BurgesRL06}, LambdaMart~\citep{burges2010ranknet,DBLP:conf/www/HuWPL19} and LambdaFM~\citep{DBLP:conf/cikm/YuanGJCYZ16}. In particular, LambdaRank is widely used as the underlying model in studies comparing the optimization of different metrics. LambdaRank-based results in~\citep{DBLP:conf/sigir/DonmezSB09, DBLP:journals/ir/YilmazR10} show that optimizing for informative metrics can lead to good performance, not only when evaluating with the same metric, but also with others. This insight invites to revisit pairwise learning recommender systems by experimenting with other metrics to optimize, even if they are not the evaluation target. Similarly, LambdaFM~\citep{DBLP:conf/cikm/YuanGJCYZ16} was deployed to assess  effectiveness with respect to three metrics, namely $nDCG$, $RR$ and $AUC$, by optimizing for $nDCG$. Optimal performance was achieved with respect to $nDCG$ and $RR$.

Another way of achieving optimal ranking with respect to a given metric is to deploy a \emph{listwise} LTR approach. This approach looks at the entire ranked list for optimization, and therefore better resembles the concept of direct metric optimization than the (indirect) pairwise LTR approach.  
Although straightforward and close in nature to LTR, listwise methods have to deal with loss functions containing integer ranking positions, which causes non-smoothness and therefore non-differentiability. A common way to deal with this problem is to approximate the indicator function by a differentiable alternative. CofiRank~\citep{DBLP:conf/nips/WeimerKLS07} was one of the first works addressing this issue by choosing to minimize the $(1 - nDCG)$ loss with a structured estimation. 
Another popular method is to use a smooth function, such as a sigmoid or ReLU~\citep{DBLP:conf/icml/NairH10}, to approximate the non-smooth indicator function. This method has been widely applied for optimizing $DCG$~\citep{DBLP:conf/bigdataconf/LiangHDH18}, $AP$~\citep{DBLP:conf/sigir/ShiKBLHO12} and $RR$~\citep{DBLP:conf/recsys/ShiKBLOH12}. 
Rather than optimizing the whole list and taking items at the bottom into account, \citet{DBLP:conf/bigdataconf/LiangHDH18} proposed Top-N-Rank, which focuses on the top ranked items and uses a listwise loss with a cutoff to directly optimize for $DCG@k$. 

Despite this rich track record of attempts to learn a ranking by metric optimization, still insufficient is known about what metric to optimize for in order to obtain the best performance according to some evaluation metric. Moreover, we believe that the scope of the metrics to consider could further be expanded to broaden the possibilities for improving ranking effectiveness beyond what has been tried so far. 
In this paper, to conduct our experimental assessment regarding ranking effectiveness for recommendation, we follow both the pairwise and listwise LTR approaches, and consider different IR metrics to optimize and assess ranking performance. Specifically, we add the $RBP$ metric as a promising candidate to the set of typically deployed $RR$, $AP$ and $nDCG$.

\section{Methods}\label{sec:method}


In this section we describe the design choices and methodology behind our experimental approach to acquire new insights into the issues related to generating recommendations through optimizing IR metrics. We start by introducing our underlying recommendation model with the notation and terminology used throughout the paper. Then, we describe the four IR metrics we choose to optimize. Finally, we define the corresponding objective functions we deploy for optimization in both the pairwise and listwise cases. In this way, we put special emphasis on the definitions of objective functions for $RBP$ that we introduce in this paper.

\subsection{Recommendation Model}\label{subsec:recmodel}

Recommender systems are meant to recommend "items" (in a general meaning of the term) to users according to their preferences. For a system with $M$ users and $N$ items, ground-truth user-item interaction data can be represented by a matrix $Y$ with dimensions $M \times N$. We consider a binary relevance scenario in this paper, which implies that elements in $Y$ are either $y_{ui}=1$, indicating a positive interaction (preference) between a user $u$ and an item $i$, or $y_{ui}=0$, indicating either a negative interaction (e.g. a `dislike'), or no interaction between $u$ and $i$. We refer to items with the positive (negative) interaction as the positive (negative) items. We assume that an arbitrary user $u$ generated $m_u$ positive interactions across all items.

Following the practice from other ranking-based recommendation approaches that target direct metric optimization~\citep{DBLP:conf/recsys/ShiKBLOH12, DBLP:conf/bigdataconf/LiangHDH18, DBLP:conf/icdm/LeeL15}, in this paper we choose Matrix Factorization (MF)~\cite{DBLP:conf/uai/RendleFGS09} as the recommendation model. Although collaborative filtering can be achieved via more advanced methods such as Neural Collaborative Filtering \cite{DBLP:conf/www/HeLZNHC17}, Collaborative Variational Autoencoders \cite{DBLP:conf/kdd/LiS17} and Graph Neural Networks \cite{DBLP:conf/sigir/0001DWLZ020, DBLP:conf/aaai/ChenWHZW20}, we still choose the base Matrix Factorization model because our aim in this paper is to study the relative merits of metrics. A more comprehensive experiment to assess generalizability with other models is left for future work.

The users and items are thus represented by latent factor matrices $U^{M \times D}$ and $V^{N \times D}$, respectively, where $D$ is the number of latent factors. Using the latent vectors of users and items, a recommendation model can predict the relevance of items for each user, and store the scores in the matrix $F^{M\times N}$, with the element $f_{ui}$ representing the predicted relevance of item $i$ to user $u$.
The ranking position $R_{ui}$ corresponding to the relevance score $f_{ui}$, is an integer ranging from 1 to $N$, calculated from a pairwise comparison between the predicted relevance score for item $i$ and all other items:
\begin{equation}
    \label{eq:rank}
     R_{ui} = 1 + \sum_{j=1\setminus i}^{N} \mathbb{I}\left(f_{uj} > f_{ui}\right)~,
\end{equation} 
\noindent where $\mathbb{I}(\cdot)$ denotes the indicator function. 

\subsection{Metrics}

As introduced before, we consider four metrics to optimize when training the ranking mechanism of a recommender system: $RR$, $AP$, $nDCG$ and $RBP$. These metrics, assessing the recommendation performance for user $u$, can be formulated as follows:
\begingroup
\allowdisplaybreaks
\begin{align}
    nDCG(u) &= \frac{DCG(u)}{iDCG(u)} \nonumber \\
    &= \frac{\sum_{i=1}^{N} \left(2^{y_{ui}}-1\right) / \log_2 (R_{ui} + 1)}
        {\sum_{i=1}^{m_u} 1 / \log_2 (i + 1)}~, \label{eq:ndcg}\\
    AP(u) &= \frac{1}{m_u}\sum_{i=1}^N \frac{y_{ui}}{{R_{ui}}} \sum_{j=1}^{N}y_{uj}\mathbb{I}(R_{uj} \leq R_{ui})~,\label{eq:ap} \\
    RR(u) &= \sum_{i=1}^N \frac{y_{ui}}{R_{ui}}
        \prod_{j=1}^N \left(1-y_{uj}\mathbb{I}(R_{uj} < R_{ui})\right)~,\label{eq:rr} \\
    RBP(u;p) &= (1-p)\sum_{i=1}^{N} y_{ui} p^{R_{ui}-1}~.\label{eq:rbp}
\end{align}
\endgroup

According to the $RBP$ formulation originally proposed by \citet{DBLP:journals/tois/MoffatZ08}, $p$ is a constant parameter ranging from 0 to 1, indicating the degree of persistence of a user. A high persistence models a user that is willing to explore items deep down the ranked list. The theoretical upper limit of $RBP$ is 1 when $N$ is infinite, which means 1 is never reached in practice~\cite{DBLP:journals/ir/SakaiK08}. To align the range of $RBP$ with other metrics used in the paper and in this way make it more comparable, we choose to optimize $nRBP$ instead, which normalizes the bare $RBP$ by the maximum obtainable with $m$ positive items:
\begin{align}
    nRBP(u;p) &= \frac{RBP(u;p)}{iRBP(u;p)} \nonumber \\
    &= \frac{\sum_{i=1}^{N} y_{ui} p^{R_{ui}-1}}{\sum_{i=1}^{m_u}p^{i-1}} = Z(p,m_u)RBP(u;p)~,
\end{align}
\noindent where $Z(p, m_u)=1/(1~-p^{m_u})$, serving as a normalization factor.


\subsection{Pairwise Metric Optimization}\label{subsec:pairwise}

Following the same rationale as in Section~\ref{subsec:recmodel}, we choose LambdaRank~\citep{DBLP:conf/nips/BurgesRL06}, the base Lambda gradient ranking model, as the pairwise LTR approach. We do not consider the more complex LambdaMART~\cite{burges2010ranknet} and LambdaFM~\cite{DBLP:conf/cikm/YuanGJCYZ16} to avoid the effect of other factors such as model ensemble and dynamic negative sampling strategy.
Derived from RankNet~\cite{DBLP:conf/icml/BurgesSRLDHH05}, LambdaRank aims at obtaining smooth gradients for optimization by calculating the performance gain from swapping the position of documents $i$ and $j$ in a ranked list.
For the $\lambda$-optimization of $RR$, $AP$ and $nDCG$, we follow the existing approaches proposed by \citet{DBLP:conf/sigir/DonmezSB09}. To the best of our knowledge, LambdaRank using $RBP$ has not been formally proposed yet, so we define the $\lambda$-optimization of $nRBP$ following the same principles.

The cost for $\lambda$-optimizing an item pair $(i, j)$ for user $u$ is
\begin{equation}
    C_{uij} = -S_{uij}o_{uij} + \ln\left(1 + e^{S_{uij}o_{uij}}\right)~,
\end{equation}
where $S_{uij}$ equals $+1$ or $-1$ depending on whether the ground truth label $y_{ui}$ is larger than that of $y_{uj}$, and the term $o_{uij} \equiv f_{ui} - f_{uj}$ represents the difference of the predicted relevance scores. 
The derivative of the cost with respect to $o_{uij}$ can be formulated as
\begin{equation}
    \frac{\delta C_{uij}}{\delta o_{uij}} = -S_{uij} + \frac{S_{uij}e^{S_{uij}o_{uij}}}{1+e^{S_{uij}o_{uij}}} = 
    -\frac{S_{uij}}{1+e^{S_{uij}o_{uij}}}~.
\end{equation}

In order to reward the positive gains and punish the negative, the $\lambda$-gradient for $nRBP$ can thus be written as
\begin{equation}
    \lambda_{uij} = S_{uij}\left|\Delta nRBP(R_{ui}, R_{uj};p) \frac{\delta C_{uij}}{\delta o_{uij}} \right|~,
\end{equation}
where $R_{ui}$ and $R_{uj}$ are the ranking positions of the item pair, calculated as in Eq.~\eqref{eq:rank}, and $\Delta nRBP(R_{ui}, R_{uj};p)$ is the difference between the corresponding nRBP values. This leads to the $RBP$-based $\lambda$-gradient formulated as
\begin{equation}
\label{eq:grad}
    \lambda_{uij}=S_{uij}\left| \frac{ Z(p,m_u)(1-p)\left(y_{ui} p^{R_{ui}-1} - y_{uj} p^{R_{uj}-1}\right) } { 1+e^{S_{uij}o_{uij}}} \right|~.
\end{equation}




\subsection{Listwise Metric Optimization}\label{subsec:listwise}

Pairwise methods, such as LambdaRank, can easily avoid the issue of non-smoothness of the optimized ranking metrics. However, and despite their success, eluding this issue in the listwise approach remains an open problem~\citep{DBLP:conf/sigir/DonmezSB09}. Successfully addressing this challenge is important because, in that way, the optimization process becomes more intuitive,
straightforward and natural \citep{DBLP:books/daglib/0027504}.
As mentioned in Section~\ref{sec:related}, methods have already been proposed for listwise optimization of $RR$, $AP$ and $nDCG$. The underlying principle is to approximate the non-differentiable indicator function with a smooth alternative. This has been done by deploying either a sigmoid function~\citep{DBLP:conf/sigir/ShiKBLHO12,DBLP:conf/recsys/ShiKBLOH12} or the Rectified Linear Unit (ReLU) \citep{DBLP:conf/icml/NairH10} as proposed in~\citep{DBLP:conf/bigdataconf/LiangHDH18}. To make a consistent comparison, in this paper we use a sigmoid for all metrics. 

Apart from the difference in the choice of the smoothing function, compared to TFMAP~\citep{DBLP:conf/sigir/ShiKBLHO12} and CLiMF~\citep{DBLP:conf/recsys/ShiKBLOH12}, Top-N-Rank~\citep{DBLP:conf/bigdataconf/LiangHDH18} is also distinctive for the way the approximated ranking position $\tilde{R}_{ui}$ is modeled. TFMAP and CLiMF model $\tilde{R}_{ui}$ using only the predicted score $f_{ui}$. In contrast, Top-N-Rank infers the predicted ranking position by looking at the whole recommendation list. It follows the idea from Eq.~\eqref{eq:rank} to get the ranking position by pairwise comparison, which is closer in nature to sorting.  

Taking all the above into account, and following the same rationale as in Sections~\ref{subsec:recmodel} and~\ref{subsec:pairwise}, we do not contemplate more complex techniques such as Boosting \citep{DBLP:conf/nips/ValizadeganJZM09, DBLP:journals/apin/GhanbariS19} or multi-agent learning \citep{DBLP:conf/cikm/ZouLAWZ19}, so that metrics are compared on a base recommendation model derived from Top-N-Rank. Specifically, we replace ReLU by a sigmoid function and approximate the ranking position as
\begin{equation}
    \label{eq:approx_rank}
    \tilde{R}_{ui}= 1 + \sum_{j=1\setminus i}^{N} \sigma(f_{uj} - f_{ui})~,
\end{equation} 
where $\sigma (x)=1 / (1+e^{-x})$.
Accordingly, the approximation of the indicator functions in Eqs.~\eqref{eq:ap} and \eqref{eq:rr} can be formulated as
\begin{align}
    \label{eq:sigmoid}
        \mathbb{I}(R_{uj} < R_{ui}) &=
        \mathbb{I}(f_{uj} > f_{ui}) \approx
        \sigma(f_{uj}-f_{ui})~,\\
         \sum_{j=1}^N\!\mathbb{I}(R_{uj}\!\leq\!R_{ui}) &=
         1\!+\!\sum_{j=1}^N\!\mathbb{I}(R_{uj}\!<\!R_{ui}) \approx
         1\!+\!\!\sum_{j=1\backslash i}^N\!\!\sigma(f_{uj}\!-\!f_{ui})~.
\end{align}



Top-N-Rank, however, optimizes metrics with a cutoff, so it does not use information from all items. As indicated by \citet{DBLP:conf/sigir/DonmezSB09}, optimizing $nDCG$ on the whole item list can lead to significantly better $nDCG@10$ performance than directly optimizing $nDCG@10$. Consequently, we choose to eliminate the cutoff.
Since the target is to maximize the approximated IR metrics, we can consider their additive inverse as the loss functions for optimization. Based on the above, the $RR$, $AP$ and $nDCG$ loss functions for user $u$ can be formulated as follows:
\begingroup
\allowdisplaybreaks
\begin{align}
    L_{nDCG}(u) &= -\frac{\sum_{i=1}^N (2^{y_{ui}}-1) / \log_2(\tilde{R}_{ui}+1)}
        {\sum_{i=1}^{m_u} 1 / \log_2 (i + 1)}~,\label{eq:loss_ndcg}\\
    L_{AP}(u) &= -\frac{1}{m_u}\sum_{i=1}^N \frac{y_{ui}}{\tilde{R}_{ui}}
    \left(1+\sum_{j=1\backslash i}^N y_{uj}\sigma(f_{uj}\!-\!f_{ui})\right)~,\label{eq:loss_ap} \\
    L_{RR}(u) &= -\sum_{i=1}^N \frac{y_{ui}}{\tilde{R}_{ui}}\prod_{j=1\backslash i}^N\left(1-y_{uj}\sigma(f_{uj}- f_{ui})\right)~.\label{eq:loss_rr}
\end{align}
\endgroup

To the best of our knowledge, no method has been proposed yet for $RBP$-based listwise optimization. Inspired by the loss definitions for the other metrics, and again choosing to work with the normalized formulation of $nRBP$, we introduce the method for defining the corresponding loss function as follows. 


By virtue of the monotonicity of the logarithm function, the recommendation model that optimizes $nRBP(u;p)$ also optimizes
\begin{equation}
    \ln\left(\frac{nRBP(u;p)}{m_u}\right) =
    \ln\left(\frac{RBP(u;p)}{m_u}\right) - \ln\left(iRBP(u;p)\right) \label{eq:lognRBP}~.
\end{equation}
Note that the second term is a constant for each user, so we focus on the first term. Based on Jensen's inequality, we can derive a lower bound for the first term as follows:
%
\begin{align}
\label{eq:lower_bound}
\begin{split}
    \ln\left(\frac{RBP(u;p)}{m_u}\right) &= \ln(1-p) + \ln\left(\frac{1}{m_u}\sum_{i=1}^N y_{ui}p^{\tilde{R}_{ui}-1} \right) \\
     &\geq \ln(1-p) + \frac{1}{m_u}\sum_{i=1}^N{\ln\left(y_{ui}p^{\tilde{R}_{ui}-1}\right)} \\
     &= \ln(1-p) + \frac{1}{m_u}\sum_{i=1}^N{y_{ui}(\tilde{R}_{ui}-1)\ln(p)}~.
     \end{split}
\end{align}
Note that the last equality holds because only $y_{ui}=1$ contributes to the summation, and the two remaining logarithmic terms are constant across users.

Because $p\in(0,1)$, $\ln(p)$ is negative, so maximizing the formulation in Eq.~\eqref{eq:lower_bound} becomes equivalent to minimizing $\sum_{i=1}^{N} y_{ui} (\tilde{R}_{ui}-1)$. In this way, our $RBP$-based optimization of the ranking treats all relevant items equally and aims at bringing them close to the top. Although convenient and intuitive, this function does not have common bounds across users. Therefore, if used alone as the optimization objective, it would make the training process sensitive to specific users. To (partially) resolve this issue, we come back to the second term in Eq.~\eqref{eq:lognRBP}, find its own ``lower bound'' using Jensen's inequality and subtract it from Eq.~\eqref{eq:lower_bound} to make the lower bound equal 0 for all users. After dropping the logarithms, the regulated $nRBP$ loss for user $u$ can now be denoted as
\begin{align}
\label{eq:loss_func}
L_{nRBP}(u)= \sum_{i=1}^{N} y_{ui} (\tilde{R}_{ui}-1) - \sum_{j=1}^{m_u} (j-1)~.
\end{align}

In this way, the optimization of $nRBP$ becomes equivalent to an elegant direct optimization for the rank position of the relevant items. In view of the fact that the ideal situation leads to ranking all relevant items at the top, the listwise loss inspired by $RBP$ shows potential for achieving high ranking effectiveness across different evaluation criteria, making it an informative metric. Furthermore, $L_{nRBP}$ is independent of $p$, which makes it possible to conduct $RBP$-based optimization for different $p$ values in one unified framework. We note here once again that the regularization in Eq.~\eqref{eq:loss_func} only has effect on the lower bound of the loss value range. The loss value remains user-sensitive and can still be arbitrarily large depending on the number of interactions $m_u$. We analyze the consequences of this in Section~\ref{subsec:ind}.

\subsection{Model Learning}

In recent years, Adaptive Moment Estimation (Adam)~\citep{DBLP:journals/corr/KingmaB14} has become one of the most popular optimizers. Compared to traditional optimizers like Stochastic Gradient Descent (SGD), its insensitivity to hyper-parameters and faster convergence makes it widely deployed in machine learning models. Despite these advantages, Adam tends to suffer from convergence and generalization power~\citep{DBLP:conf/nips/ZhouF0XHE20, DBLP:journals/tois/MoffatBST17}. Because in this paper we investigate the generalization power of optimizing for different metrics, 
we still choose to optimize all our models with SGD.

\section{Experimental Design}\label{sec:exp}

Our goal is to investigate the capabilities of metrics used for optimization when the recommendation performance is assessed by the same or other metrics. In the following we explain the selection of datasets and experimental protocol for our experiments.

\subsection{Datasets}

We selected four widely-used and real-world datasets to experiment with a diverse set of data. Two of them, CiteULike-a~\citep{DBLP:conf/ijcai/WangCL13} and Epinions~\citep{DBLP:conf/wsdm/TangGL12}, contain unary data, while the other two, Sports \& Outdoors and Home \& Kitchen, are datasets with graded ratings from Amazon~\citep{DBLP:conf/emnlp/NiLM19}. Amazon datasets contain integer relevance scores ranging from 1 to 5, so we need to binarize them before they can be used by our LTR methods. We choose to consider as positive only ratings of 4 and 5, which surely reflect a positive preference, and every other rating as a negative preference. In addition, and as is common in experimentation with recommenders, the absence of a rating is also taken as a negative interaction in all datasets \citep{DBLP:conf/recsys/SunY00Q0G20}.
As shown by \citet{DBLP:journals/ir/CanamaresCM20}, although users with few ratings might exist in commercial services, they are usually filtered out in offline experiments because the lack of data leads to unreliable performance measurements. To address this issue, we only keep users with at least 25 relevant interactions in all datasets. The post-processed dataset statistics are shown in Table~\ref{tab:dataset}.

\subsection{Experimental Protocol}\label{subsec:protocol}

We use LensKit \citep{DBLP:conf/cikm/Ekstrand20} to randomly split the data into training and test sets , stratifying by user: we sample 80\% of their interactions for training, and leave the rest for testing. As a result, each user has at least 20 relevant interactions in the training set, and at least 5 in the test set. The evaluation metrics used are $RR$, $AP$, $nDCG$ and $RBP$ with $p$ equal to $0.8$, $0.9$ and $0.95$, so that we can assess recommendation performance under different degrees of user persistence.
For pairwise optimization using LambdaRank, these six metrics are in line with six separate loss functions. In the listwise context, however, we have in total 4 loss functions because the loss for $nRBP$ is independent of $p$. All models are trained with a considerable number of epochs (3,000), so that every metric has a more than reasonable chance to achieve its best performance. For each model, we select the epoch yielding the best performance on the corresponding evaluation metric.
For the optimization of listwise $nRBP$, the optimal epoch is chosen for each of the 3 values of $p$ separately, so that we actually have 3 different models.

To reduce random error due to data splitting, we adopt a Monte Carlo cross-validation approach~\cite{dubitzky2007fundamentals} and create three independent splits per dataset. 
\citet{DBLP:conf/sigir/ShiKBLHO12} indicate that, for IR metrics that only rely on the ranking positions of relevant items, there is no need to consider all irrelevant items when training. As a result, a negative sampling process can significantly speed up training without hurting the overall performance. Negative sampling is, however, not only beneficial for efficiency. According to \citet{DBLP:conf/recsys/CanamaresC20}, removing some (or even a significant number of) negative
items from both the training and test sets can make the evaluation more informative and less biased with, balancing popularity and the average relevance of items across users. Such a strategy can therefore also make the evaluation more effective. 
We choose to inform the negative sampling process by the number of relevant items for each user, so we sample as many negative items as positives, twice as many, or five times as many; we denote this as the Negative Sampling Ratio (NSR). These negative items, along with the 80\% of positives, form the complete training set for a user. In order to align the distributions of training and test sets, we complete the test set using the same approach: the remaining 20\% of positives, plus 100\%, 200\% or 500\% as many negatives.

In order to maximize the performance of the assessed models, we fine-tuned the learning rate of SGD and performed a search in the range \{0.001, 0.01, 0.1\} for LambdaRank and \{0.001, 0.01, 0.1, 1, 3, 10\} for listwise models. We also conducted a preliminary exploration on the number of latent factors for Matrix Factorization within the range \{8, 16, 32, 64, 128\}. The results showed that, although there is a positive correlation between the ranking effectiveness and the latent space dimensionality, such a correlation has no impact on the relative performance of different losses. Therefore, we do not analyze the effect of dimensionality in this paper and simply fix the number of latent factors at 32 throughout the experiments.


We implement all models in PyTorch \citep{paszke2017automatic}. To accelerate the training process, we use CUDA and CuDNN on an NVIDIA GeForce GTX 1080Ti GPU.

\begin{table}[t]
  \caption{Dataset statistics.}
  \label{tab:dataset}
  \centering{\small
  \begin{tabular}{@{}ccrrrr@{}}
    \hline
    Dataset & \#users & \#items & \#ratings & Density\\
    \hline
    CiteULike-a & 2,465	& 16,702 & 157,527 & 0.383\%\\
    Epinions & 4,690 & 32,592 &	325,154 & 0.213\%\\
    Sports \& Outdoors & 9,123 & 119,404 & 342,311 & 0.031\% \\
    Home \& Kitchen & 20,531 & 222,472 & 795,845 & 0.017\% \\ \hline
  \end{tabular}}
\end{table}

\section{Results}\label{sec:res}

\begin{figure*}[!t]
    \centering\includegraphics[width=.5\textwidth]{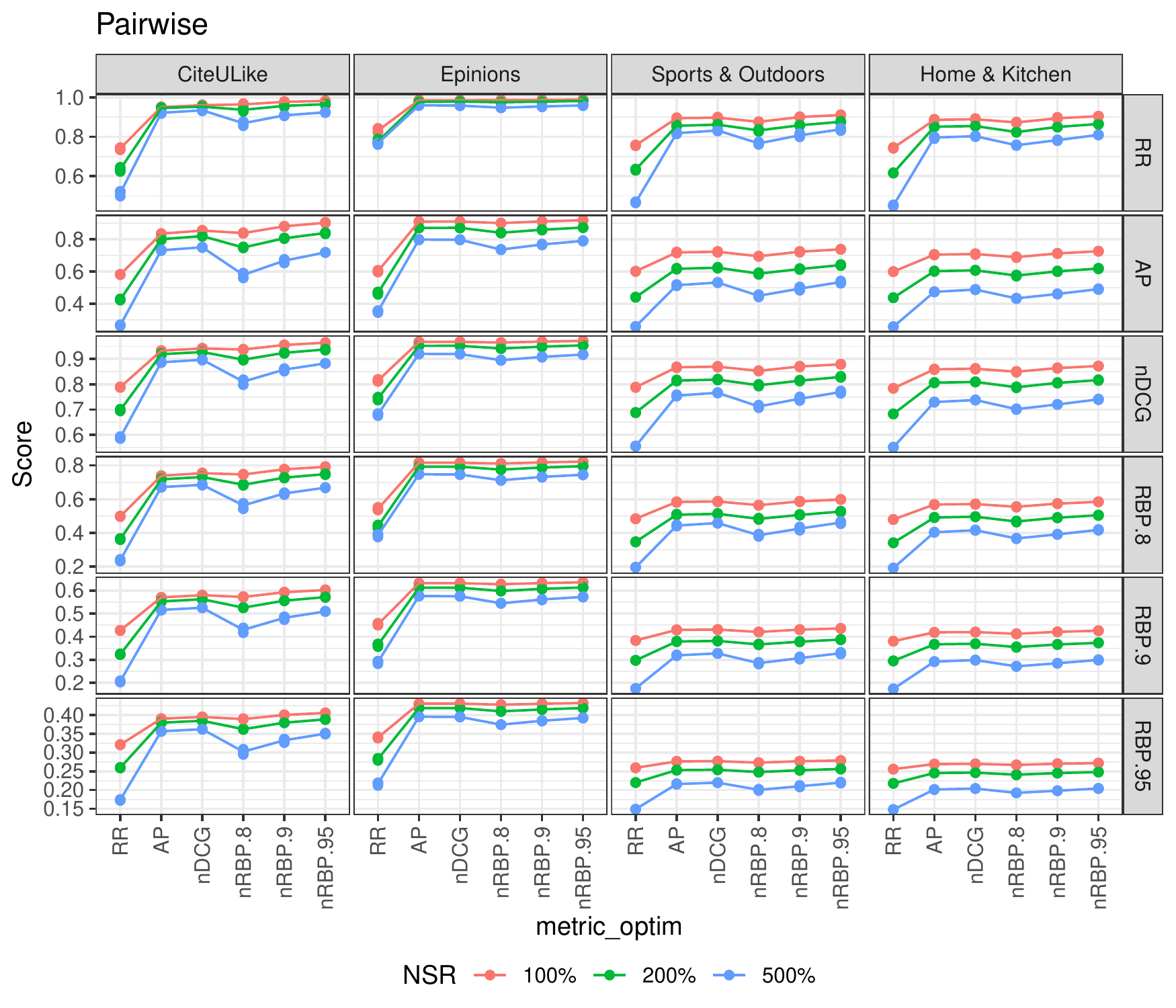}\includegraphics[width=.5\textwidth]{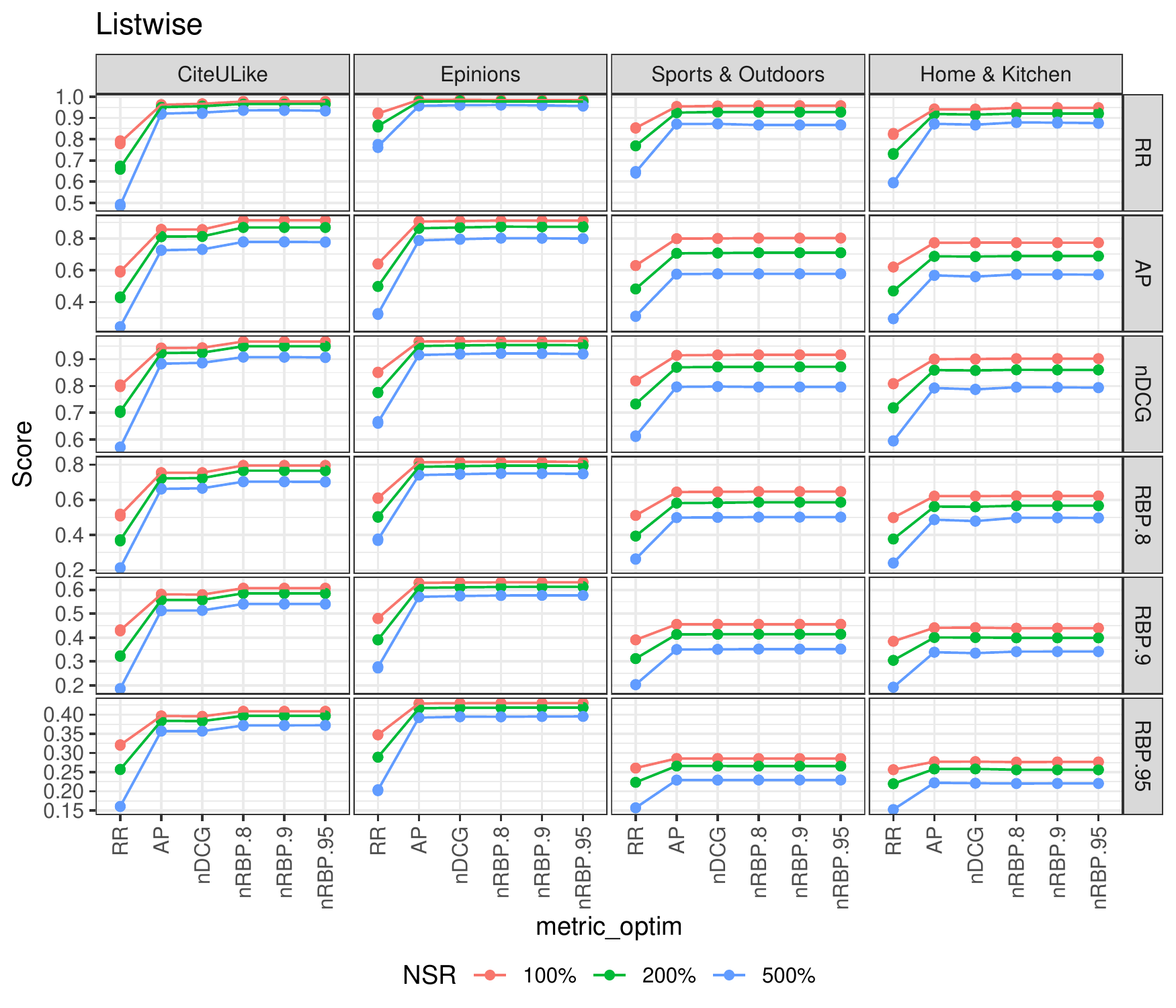}
    \caption{Overall performance of pairwise and listwise methods. \texttt{metric\_optim} denotes the metric-based losses for optimization, and a panel row denotes an evaluation metric. All results are averaged over the 3 data splits. Note that y-axes vary per row.}\label{fig:performance}
\end{figure*}

In this section we compare the effectiveness of different metrics when used for optimization in ranking-based recommender. The goal of the analysis is threefold. First, in Section~\ref{subsec:res} we focus on the overall performance of pairwise and listwise models and investigate whether the practice of optimizing for the metric used in evaluation is justified in ranking-based recommendation. Then, in Section~\ref{subsec:rbp} we conduct a deeper analysis on the impact of a metric chosen for optimization on the ranking effectiveness assessed by different evaluation metrics. In doing so, we especially focus on the performance of the $RBP$-inspired objective functions introduced in this paper. Finally, in Section~\ref{subsec:ind} we investigate the effect of different metric optimization strategies on the recommendation utility for active and inactive users, with special emphasis on the impact of the missing upper bound of the $RBP$-based listwise loss function proposed in Eq.~\eqref{eq:loss_func}.

Due to space constraints, we do not report all the results obtained in our experiments. The reported results are, however, fully representative of the complete set of results that led to the final observations, conclusions and recommendations for future work.\footnote{All data, code and full results are available at\\ \url{https://github.com/roger-zhe-li/sigir21-newinsights}.}


\subsection{Should we Optimize the Metric Used to Evaluate?}\label{subsec:res}

\begin{figure*}[!t]
    \centering\includegraphics[width=.5\textwidth]{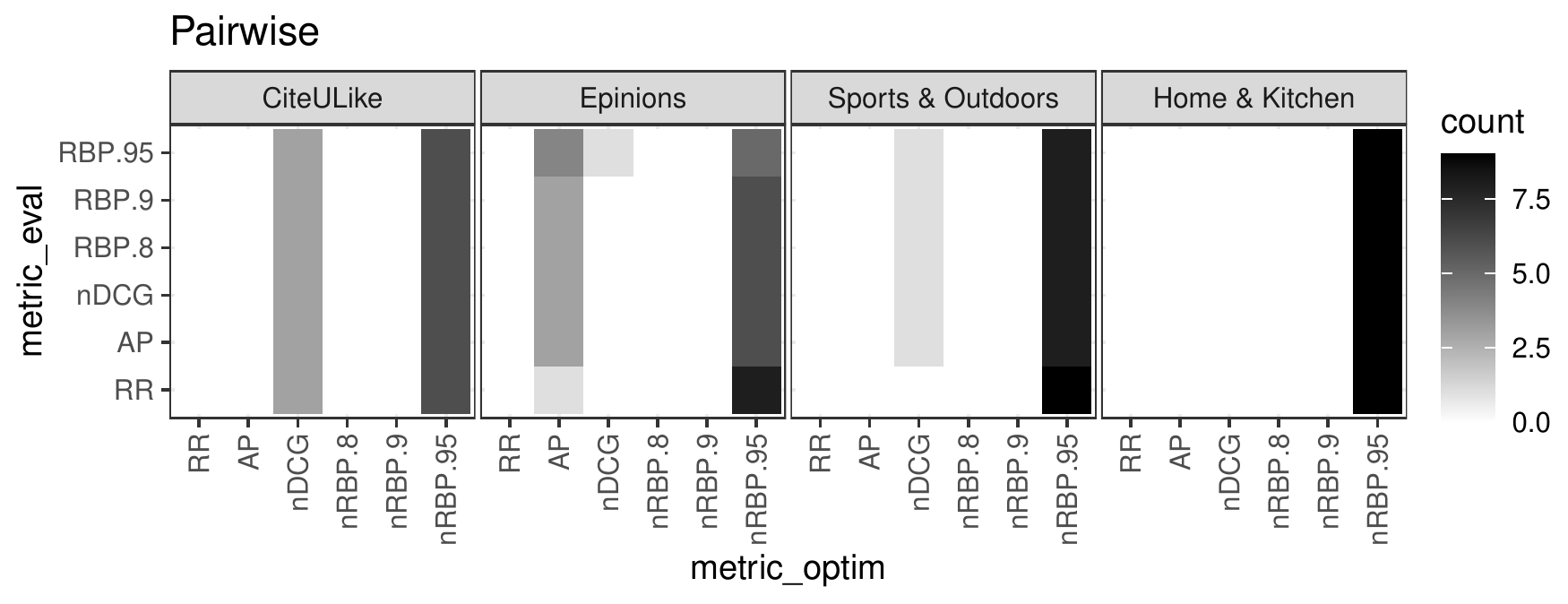}\includegraphics[width=.5\textwidth]{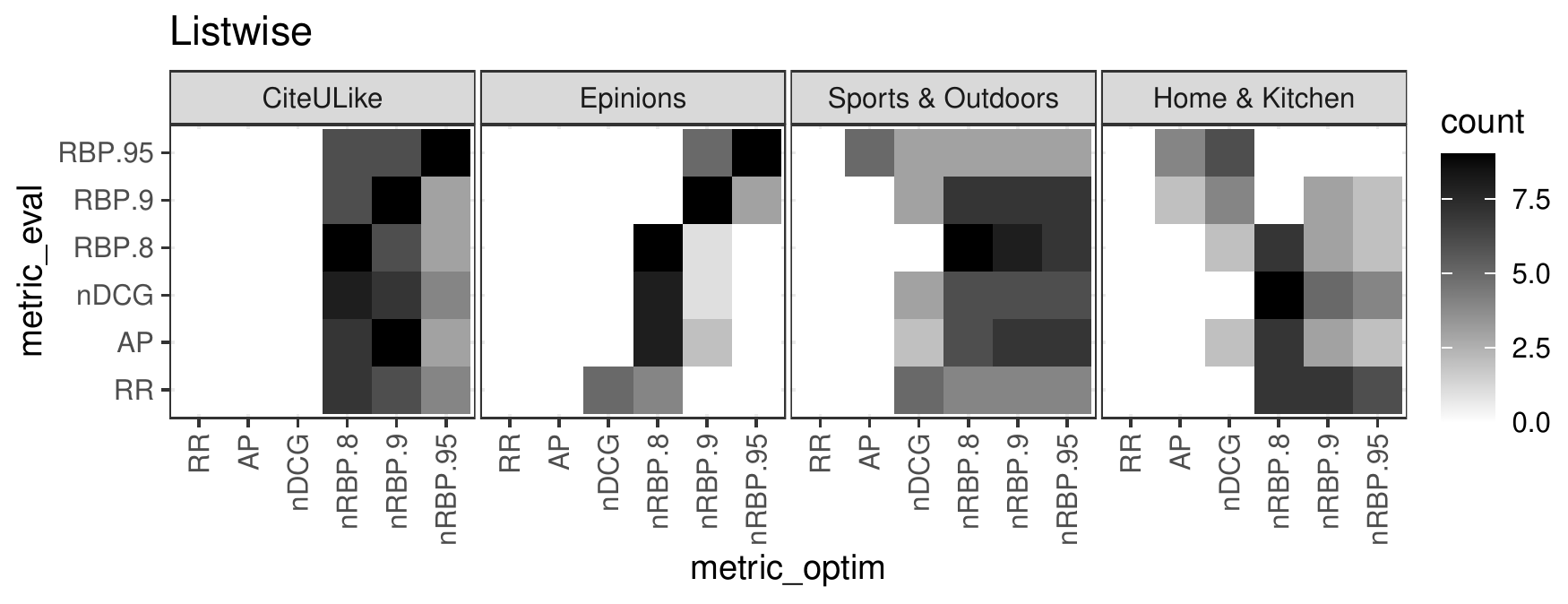}
    \caption{The frequency for each metric-based loss (\texttt{metric\_optim}) of achieving the best performance on specific evaluation metrics (\texttt{metric\_eval}), in all 3 data splits and all 3 NSR's.}\label{fig:count}
\end{figure*}

Fig.~\ref{fig:performance} shows the performance of all pairwise and listwise learning objectives on all 6 evaluation metrics. While a negative correlation can be observed between recommendation effectiveness and the NSR, this does not necessarily mean that the models trained with more irrelevant items are worse. Because the training and test sets follow the same distribution, more negative items in the test set just make the relevance prediction task harder.

Several observations can be made from this figure. First, in both pairwise and listwise models, optimizing $RR$ consistently yields significantly worse performance, even when the evaluation target is also $RR$. 
This observation supports previous findings that optimizing other metrics can achieve higher $RR$ test scores than optimizing $RR$ itself~\citep{DBLP:conf/sigir/DonmezSB09}. The explanation for this is two-fold. First, $RR$ does not exploit all the information in the training data because it only focuses on the first relevant item, resulting in suboptimal models compared to optimizing other metrics. Second, as indicated by~\citet{DBLP:conf/sigir/WebberMZS08}, $RR$ is not well correlated with other informative metrics, such as $nDCG$ and $AP$, which may optimize for $RR$ but not the other way around.
Because the performance gap of optimizing $RR$ is stable and significant, we do not include it in further analysis and instead focus on the other 5 learning objectives (\texttt{metric\_optim}), but still with all 6 evaluation metrics.

Second, listwise and pairwise methods behave differently when optimizing $nRBP$. In the pairwise context, the recommendation performance obtained by optimizing different $nRBP$ losses is varied. Specifically, we find that optimizing with $p=0.95$ outperforms optimizing with $p=0.8$ or $p=0.9$. This finding is consistent with prior research showing that $RBP.95$ is better correlated with informative metrics than with other different $p$'s~\citep{DBLP:journals/ir/SakaiK08, DBLP:conf/adcs/MoffatST12}. Such an inner-$RBP$ advantage can also be explained by the nature of the metric. Because $p$ models user persistence, a high value takes relevant items lying deeper in the list into account during training. This allows the model to account for more information, which benefits its performance. Although higher $p$ also brings a slower weight decay, which does not favor prediction for the top of a recommendation list, this effect is insignificant for systems with binary relevance, where we do not need to put highly relevant items ahead of moderately relevant ones.
In the listwise context, however, the results obtained when optimizing for $nRBP$ with different $p$ values are in general homogeneous. This shows that our $p$-independent $nRBP$ loss provides a generic method to optimize for multiple $RBP$ metrics. In this way, the concern of choosing a specific $p$ for optimization is addressed in a simple and effective way.

Last but not least, in both listwise and pairwise paradigms optimizing $nDCG$, $AP$ and $RBP$-inspired losses achieves similar stable performance across different datasets, NSRs and evaluation targets. 
Such an observation, combined with the finding that optimizing $RR$ leads to the worst ranking effectiveness when evaluating with $RR$, suggests that the practice to optimize and evaluate recommender systems with the same metric is not necessarily the best. A more detailed analysis on the relative advantages of individual metrics is given in the next section.

\subsection{Is $RBP$ More Effective as Optimization Metric than Others?}\label{subsec:rbp}




Fig.~\ref{fig:count} shows how often each metric achieved the best test performance, across evaluation metrics, when used for optimization. Since frequencies are counted from 3 splits and 3 NSR's for each evaluation metric, the frequency is expected to have a row-wise sum of 9.  
Of course, it is possible for different losses to tie and obtain the best result for a certain case, so the row-wise sum is actually larger than 9 for several evaluation metrics, specially in the listwise case. Overall, we find that in both pairwise and listwise scenarios most of the best performance cases are achieved when optimizing $RBP$-inspired metrics. More importantly, in LambdaRank we even find that optimizing $nRBP.95$ shows a significant and clear advantage over all the other metrics.

\begin{figure}
    \centering\includegraphics[scale=.51]{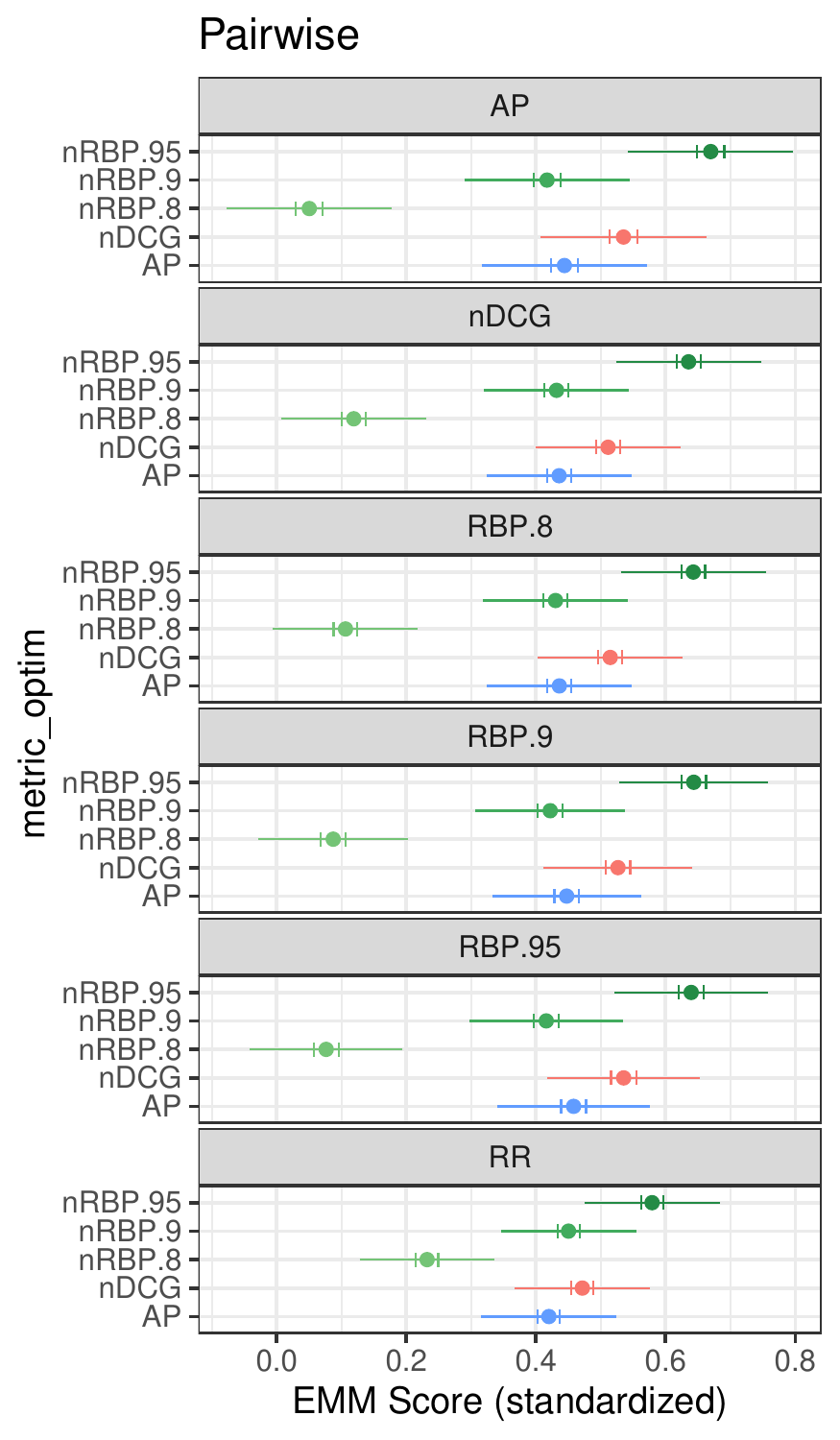}\includegraphics[scale=.51]{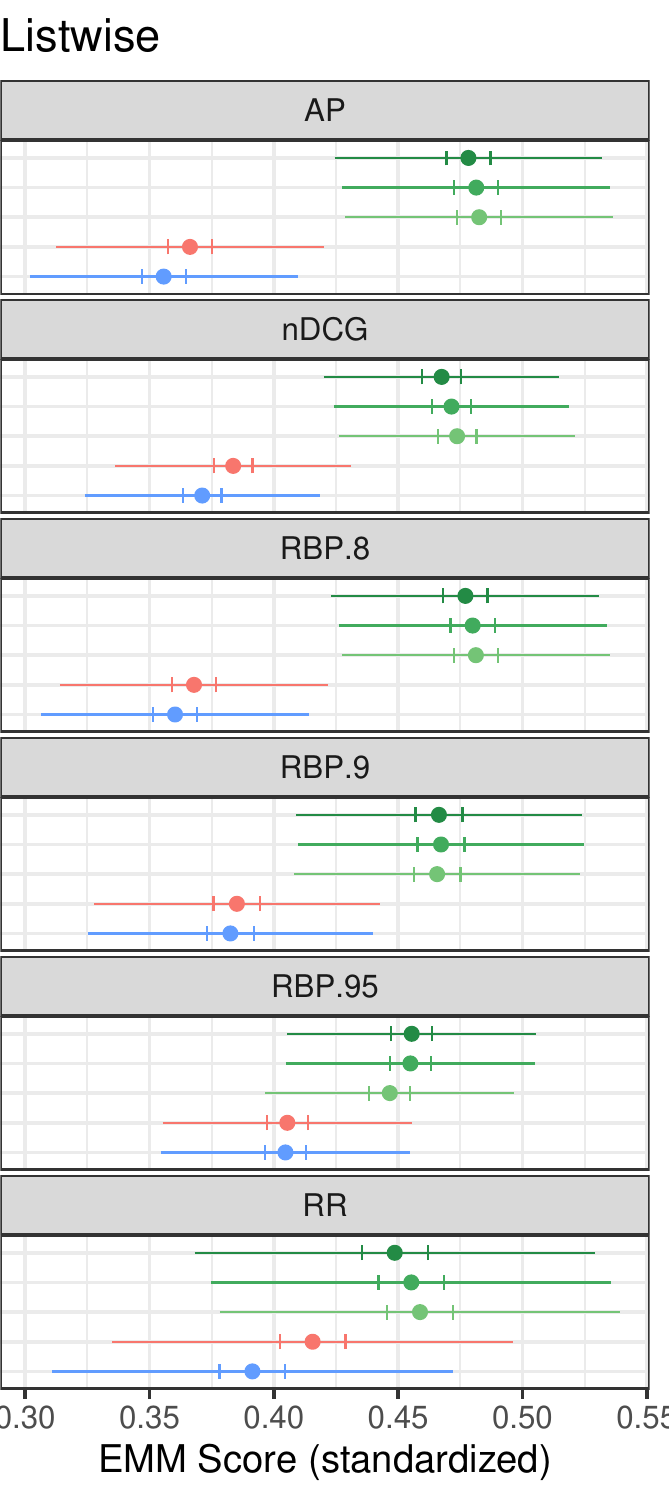}
    \caption{Estimated Marginal Mean (standardized) score for each \texttt{metric\_optim}. Each panel represents an evaluation metric. Small bounded segments represent 95\% confidence intervals. Long unbounded segments are prediction intervals.}\label{fig:stat1}
\end{figure}

\begin{figure*}
    \centering\includegraphics[width=.5\textwidth]{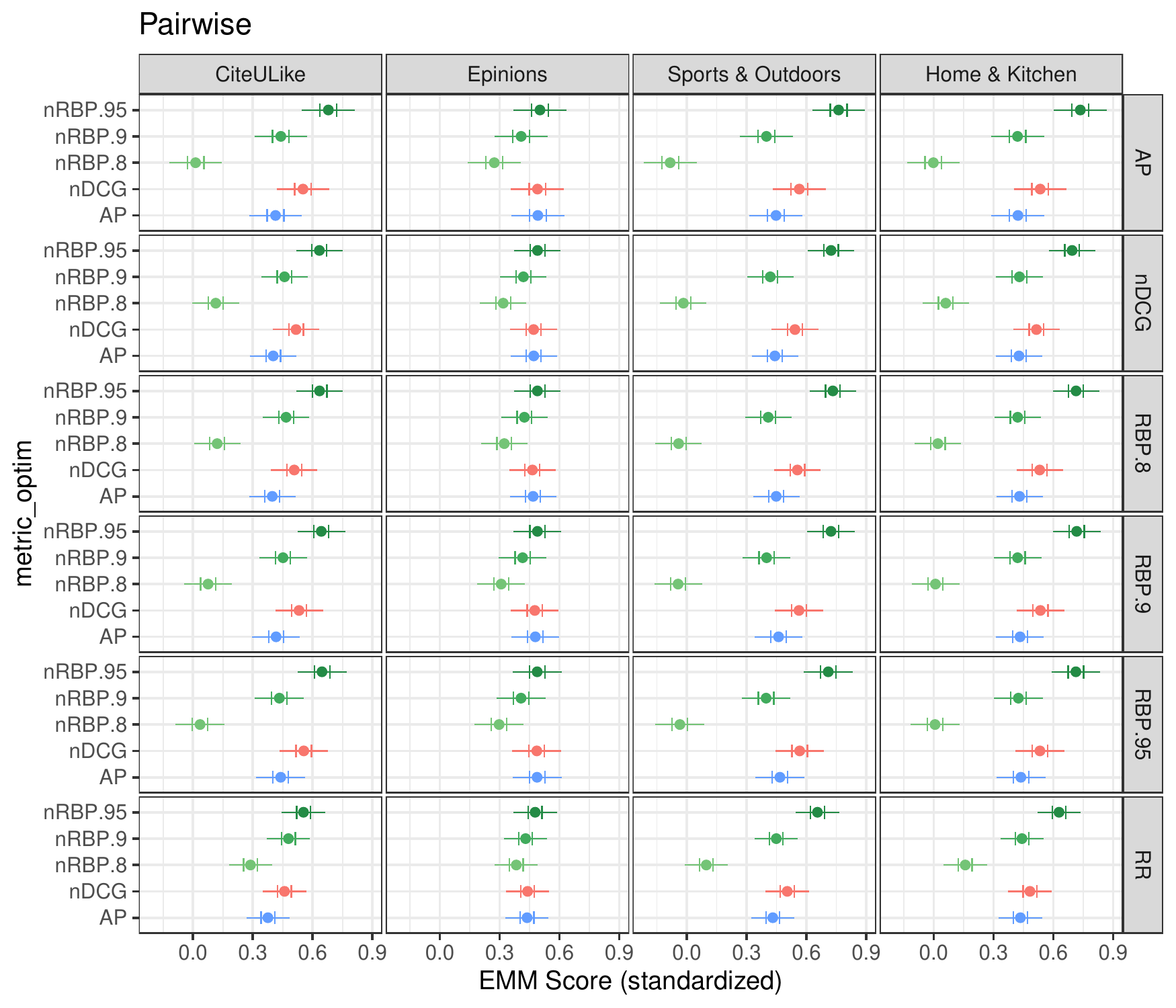}\includegraphics[width=.5\textwidth]{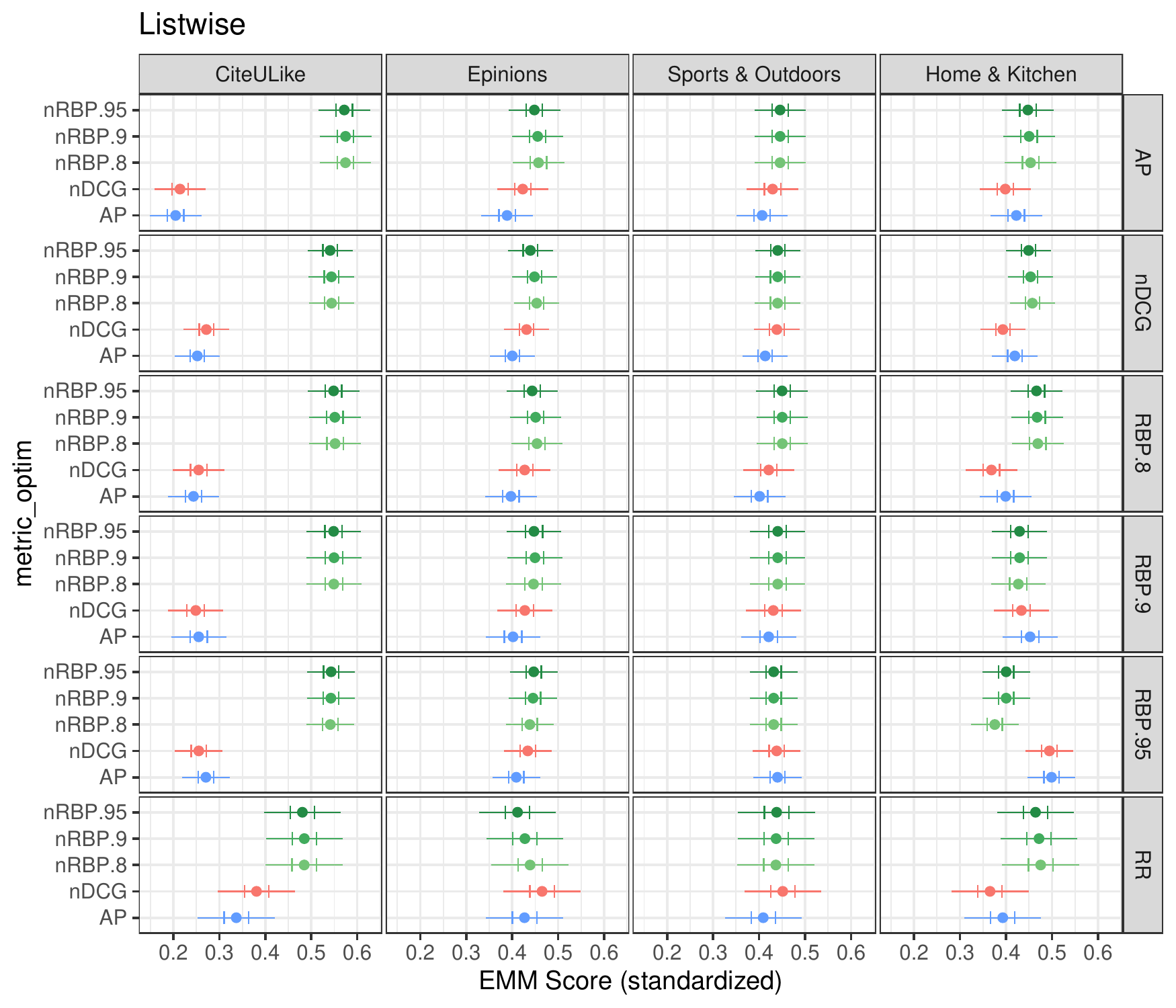}
    \caption{Estimated Marginal Mean (standardized) score for each \texttt{metric\_optim}, by evaluation metric (rows) and dataset (columns). Small bounded segments represent 95\% confidence intervals. Long unbounded segments are prediction intervals.}\label{fig:stat}
\end{figure*}

Even though $RBP$-based losses seem to achieve the best performance, Fig.~\ref{fig:performance} suggests that differences could be too small or indistinguishable from random error, so next we proceed to a statistical analysis.
First, we standardize performance scores within dataset-NSR combination to avoid homoscedasticity, because Fig.~\ref{fig:performance} evidences very different scales per dataset and NSR. This way, scores are comparable across metrics. We then fit a linear model on the standardized scores, using as independent variables the loss function, dataset and NSR, as well as their 2-factor interactions with loss function.\footnote{The inclusion of dataset and NSR main effects does not inform the model in any way because of the standardization, but we keep them to follow the hierarchy principle of linear models.} To properly compare the effect of each loss function while controlling for the other factors, we compute their Estimated Marginal Means (EMM)~\citep{emmeans}, as well as 95\% confidence and prediction intervals, that is, what to expect \emph{on average} over multiple training runs, and what to expect of an \emph{individual} training run.

Fig.~\ref{fig:stat1} presents the EMM standardized performance scores of all metric-based losses except $RR$. These overall results show that, in LambdaRank, optimizing $nRBP.95$ achieves the best performance across all 6 evaluation metrics and shows a consistent and statistically significant advantage over the other losses. In listwise models, our $RBP$-inspired loss also achieves a statistically significant advantage over the others on all 6 evaluation metrics. These observations demonstrate the power of optimizing $nRBP$.

To further explore the stability of such performance gain across different datasets, we show in Fig.~\ref{fig:stat} the EMM scores, but faceted by dataset. 
Although the general superiority of $RBP$-based learning objectives is statistically significant on average, the overlapping prediction intervals indicate that it is not always necessarily the best option.
In LambdaRank, the $nRBP.95$ loss performs best in all datasets except Epinions, where it is not statistically different from the $nDCG$ and $AP$ losses. Interestingly, we see that optimizing $nRBP.8$ lies in the opposite extreme and consistently achieves the worst performance. Even its prediction interval seldom overlaps with that of optimizing $nRBP.9$ or $nRBP.95.$
In listwise recommendation, the $RBP$-inspired loss shows significant superiority over $AP$ in a nearly consistent fashion, except for some notable cases like evaluating $RR$ in the Epinions dataset or $RBP.95$ in the Home \& Kitchen dataset. It also performs significantly better than $nDCG$ in general, but both losses yield otherwise similar performance in several cases, especially in the Epinions and Sports \& Outdoors datasets. On the CiteULike dataset, however, the advantage of $RBP$-based models is very clear and there is even no overlap between the prediction intervals, except when evaluating $RR$.

All in all, we draw the conclusion that optimizing $nRBP.95$ can help achieve recommendation effectiveness at least not worse than when optimizing other informative metrics, such as $nDCG$ and $AP$. Furthermore, the performance of optimizing $nRBP$ on different metrics is homogeneous, which means that our $RBP$-inspired listwise loss is able to help maximize $RBP$ scores regardless of $p$. Moreover, according to the unreported results from our training logs, 
this homogeneity is not only achieved in evaluation scores, but also in the convergence process. When training with listwise $RBP$-inspired losses, we manage to get the optimal $RBP$ values on all three $p$ values at a similar stage.
In some models, the epochs to get all three optimal $RBP$ scores are even the same, which means that the epoch with the optimal $RBP.95$ score also provides good scores on $RBP.8$ and $RBP.9$, and that we have the possibility to validate on only one $p$ value to satisfy different needs expressed by different values of $p$. Hence, our listwise $nRBP$ optimization can serve as a generic choice for rank-based recommendation. We can explain this by the nature of our listwise $nRBP$ loss. With our transformation in Eq.~\eqref{eq:loss_func}, we do not assign different weights to items ranked at different positions. Instead, we aim at bringing all relevant items to the top of the list, and treat all positive items as equally important. This provides the model with more abundant information to train.


The observations above point to the conclusion that, although the superiority brought by optimizing for $RBP$-based losses is not always significant and not fully consistent across datasets, we are still provided with a promising alternative metric to optimize in rank-based recommender systems. By optimizing $RBP$-based losses, we are able to get at least comparable performance as optimizing $nDCG$ and $AP$, with very clear benefits in many cases. In the following section, we will conduct an analysis seeking the source of performance advantage of $RBP$-based listwise losses.

\subsection{When to Deploy $RBP$ for Recommendation?}\label{subsec:ind}

\begin{figure*}
    \centering
    \includegraphics[width=.495\textwidth]{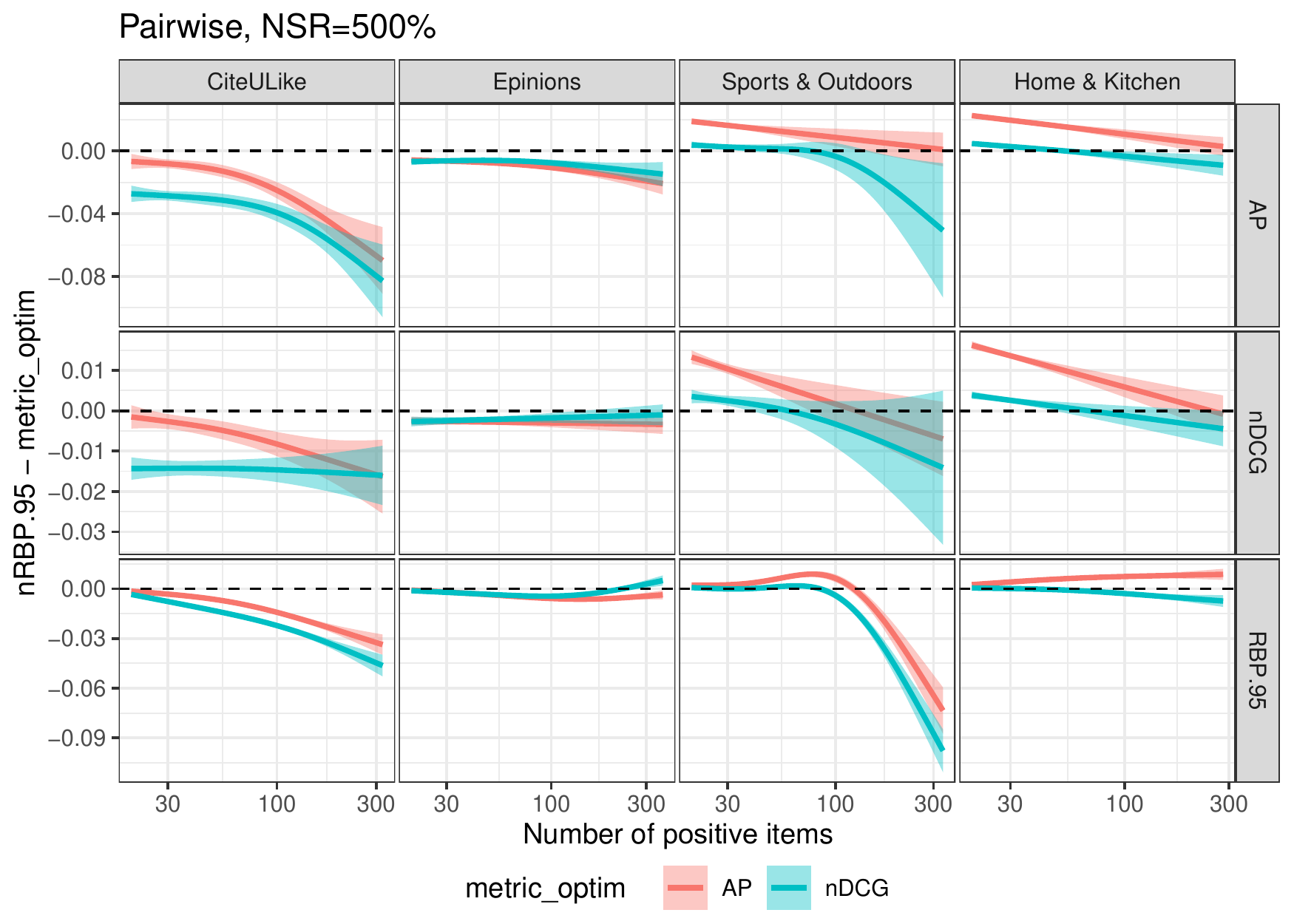}
    \includegraphics[width=.495\textwidth]{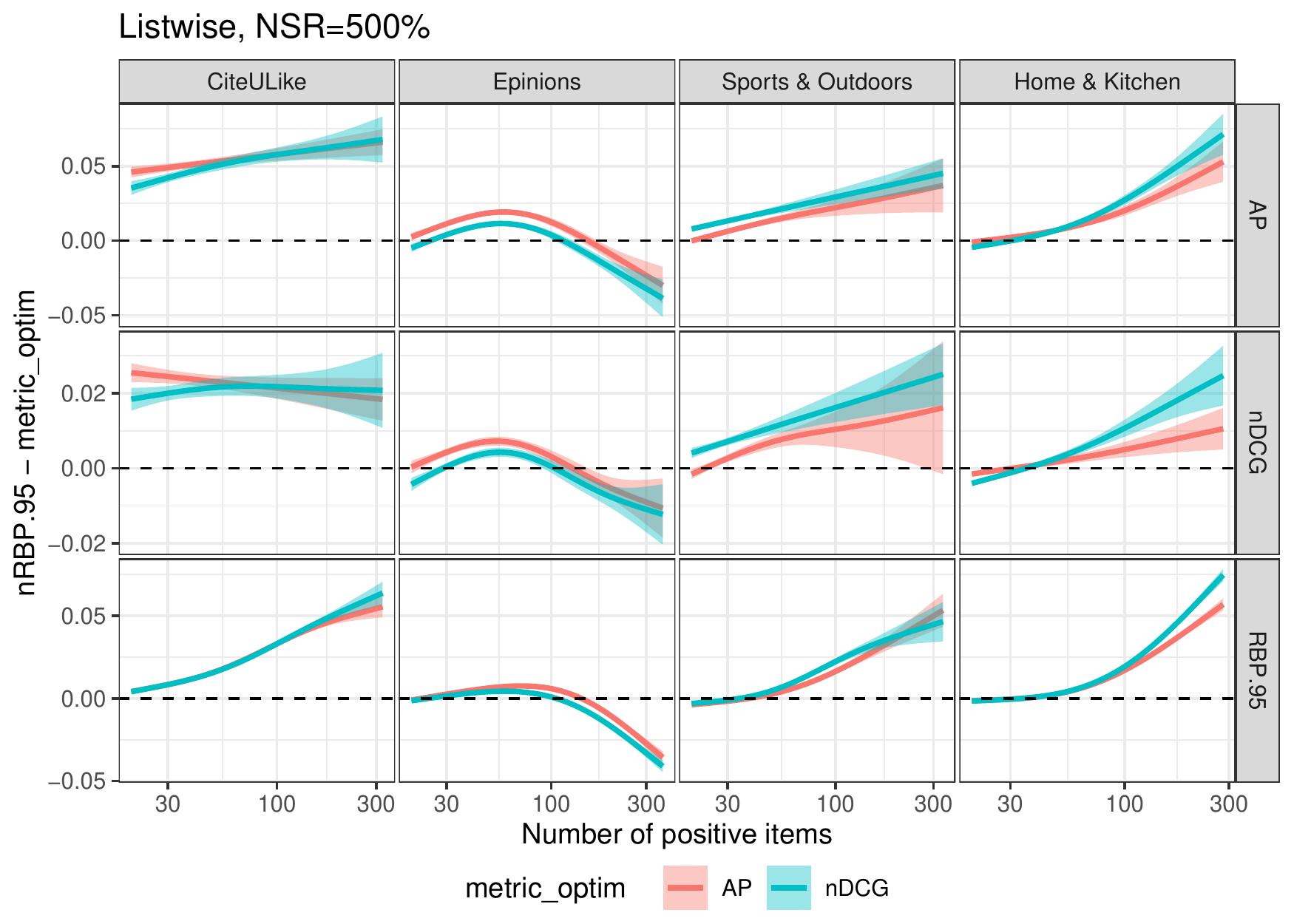}
    \caption{Score difference between optimizing $nRBP.95$ and $AP$ or $nDCG$ (higher is better for $nRBP.95$), when NSR=500\%, as a function of the number of positive items for the user. Curves show spline-smoothed fits with 95\% confidence intervals.}
    \label{fig:ind}
\end{figure*}

The analysis in the previous section indicated that optimizing $nRBP$ can provide recommendation performance comparable to that of $nDCG$ and $AP$ or even better. We may trace back the source of such effectiveness and identify the best way to deploy it if we analyze the properties of the $RBP$-based objective functions in the pairwise and listwise context. Similarly to the $nDCG$ and $AP$ losses, the pairwise $nRBP$ loss in Eq.~\eqref{eq:grad} guarantees strict bounds for the swap loss for each user, so that users are treated equally regardless of how many interactions they have. Contrary to this, the loss used for listwise $nRBP$ in Eq.~\eqref{eq:loss_func} is directly related to the predicted rank positions, which makes it have a different upper bound across users. Active users with more positive interactions are more likely to have larger loss values during training, specially for large $NSR$. Such an imbalance may benefit active users, but perhaps at the cost of sacrificing the utility for inactive users. We investigated whether this effect indeed occurs in our experiments.

Fig.~\ref{fig:ind} shows the performance difference between the $nRBP.95$ loss and $AP$ or $nDCG$ loss for every user, as a function of the number of positive items they contain in the training set. We deliberately choose to compare $nRBP.95$ because it achieves overall best results, and $AP$ and $nDCG$ for being the two next best losses, also with a properly bounded loss function. In addition, we only show results here for $NSR=500\%$, as it is expected to amplify the aforementioned bias, if any.
We find that the $RBP$-inspired losses indeed perform differently under pairwise and listwise environments. In LambdaRank, where the $nRBP.95$ loss is also strictly bounded, we do not observe a clearly positive or negative correlation between the performance difference and the user activity level. In several cases, for active users with more items available for training, optimizing $nDCG$ and $AP$ is even more advantageous than optimizing $nRBP.95$. This means that active users did not benefit from the fully bounded $nRBP$ loss, and there is no effectiveness imbalance between active and inactive users.

However, this observation does not hold in listwise models, as evidenced by the clear and significant positive correlation in all datasets except Epinions: active users do indeed benefit the most.
More interestingly, this benefit for active users is not in detriment of the inactive users. As we can see in the figure, compared to optimizing $nDCG$ and $AP$, optimizing the listwise $nRBP.95$ loss benefits all users on the CiteULike dataset, and achieves similar effectiveness for inactive users on the other 3 datasets. 
This superiority confirms our assumption that active users, whose listwise $nRBP$ loss magnitudes are larger than for other users, can indeed get more effective recommendations due to a training process biased towards their utility, without negatively influencing the less active users. Such an insight is potentially interesting for some business applications of recommender systems, where it is beneficial to maximally serve loyal users without losing the stickiness of less active users.

\section{Conclusion}\label{sec:conclusion}

Direct optimization of IR metrics has long been a hotspot in the research on ranking-based recommender systems. The intuitive and logical common practice is to build models by optimizing the same metric that will be used for evaluation. In this paper, we reported the results of an extensive experimental study aiming at acquiring new insights about the strength of the foundations behind this practice and at learning more about what metric to optimize in order to maximize recommendation effectiveness. For this purpose, we expanded the scope of metrics usually deployed to define the objective functions for LTR approaches and focused on $RBP$ as a promising alternative to other metrics such as $AP$, $nDCG$ and $RR$.

Experimental evidence on both pairwise and listwise frameworks show that optimizing $AP$, $nDCG$ and $nRBP$ generally outperforms optimizing $RR$, and that optimizing $nRBP$ is generally no less effective than optimizing $nDCG$ or $AP$. These findings challenge the practice to optimize and evaluate ranking-based recommender systems using the same metric. Furthermore, the new generic listwise $RBP$-inspired loss proposed in this paper was shown to be able to achieve the optimal performance for different values of the user persistence parameter, without the need to specify this parameter explicitly. Optimizing this loss even significantly outperformed the direct optimization of $nDCG$ and $AP$ in some cases, showing the high potential of $RBP$ for developing ranking-based recommender systems. Finally, and due to the lack of a common upper bound across users, our proposed listwise $nRBP$ loss benefits active users more than $nDCG$ and $AP$, but without hurting the effectiveness for inactive users. This makes optimization of the proposed $RBP$-based listwise loss interesting for some business application cases favoring loyal users.


For future work, we will experiment with more advanced recommendation models and larger datasets to study the extent to which our conclusions and insights generalize to other settings.
We will also analyze to what extent the exclusion of very inactive users affects our conclusions, especially with regards to the bounds of $nRBP$ losses.
Furthermore, it would be interesting to theoretically investigate the source of the effectiveness of $RBP$ even deeper.
Our results show that it is a promising metric to optimize when learning to rank for recommendation, pointing to the possibility of finding even more IR metrics that could show similar potential.
\balance
\bibliographystyle{ACM-Reference-Format}
\bibliography{sample-base}

\end{document}